\documentclass[submission,copyright,creativecommons]{eptcs}
\providecommand{\event}{SynCoP 2014} 
\usepackage{breakurl}             
\usepackage{graphicx}
\usepackage{amssymb}
\usepackage[strings]{underscore}
\usepackage[font=scriptsize,labelfont={bf}]{caption}
\usepackage[justification=raggedright,nearskip=10pt,farskip=0pt]{subfig}
\RequirePackage[log]{snapshot}

\title{Setting Parameters for Biological Models With ANIMO}
\author{Stefano Schivo
\institute{Formal Methods and Tools\\
Faculty of EEMCS\\
University of Twente\\
Enschede, The Netherlands}
\email{s.schivo@utwente.nl}
\and
Jetse Scholma \quad Marcel Karperien \quad Janine N. Post
\institute{Developmental BioEngineering\\
MIRA Institute for Biomedical Technology and Technical Medicine\\
University of Twente\\
Enschede, The Netherlands}
\email{\{j.scholma, h.b.j.karperien, j.n.post\}@utwente.nl}
\and
Jaco van de Pol \quad Rom Langerak\thanks{Corresponding author}
\institute{Formal Methods and Tools\\
Faculty of EEMCS\\
University of Twente\\
Enschede, The Netherlands}
\email{\{j.c.vandepol, r.langerak\}@utwente.nl}
}
\def\titlerunning{Setting Parameters for Biological Models With ANIMO}
\def\authorrunning{S. Schivo \emph{et al.}}

\def\tas{Timed Automata}
\begin{document}
\maketitle

\begin{abstract}
ANIMO (Analysis of Networks with Interactive MOdeling) is a software for modeling biological networks, such as e.g. signaling, metabolic or gene networks.
An ANIMO model is essentially the sum of a network topology and a number of interaction parameters. The topology describes the interactions between biological entities in form of a graph, while the parameters determine the speed of occurrence of such interactions.

When a mismatch is observed between the behavior of an ANIMO model and experimental data, we want to update the model so that it explains the new data. In general, the topology of a model can be expanded with new (known or hypothetical) nodes, and enables it to match experimental data. However, the unrestrained addition of new parts to a model causes two problems: models can become too complex too fast, to the point of being intractable, and too many parts marked as "hypothetical" or "not known" make a model unrealistic. Even if changing the topology is normally the easier task, these problems push us to try a better parameter fit as a first step, and resort to modifying the model topology only as a last resource.

In this paper we show the support added in ANIMO to ease the task of expanding the knowledge on biological networks, concentrating in particular on the parameter settings.
\end{abstract}

\section{Introduction}
The investigation of biological processes relies on computational support on a daily basis.
This happens not only because of the extremely large amount of data generated in the \emph{-omics era},
but also because many processes are simply too complex to be understood without appropriate modeling tools.
For this reason, systems biology~\cite{priami-alg-sys-bio} has become more and more important in the last several years.

A single biological network such as a signaling or gene network, may involve up to hundreds of
different molecular species. As it would be very difficult to understand the dynamic behavior of such
networks just by looking at their static representations, many tools were built~\cite{biopepa-interface,
blenx, cell-illustrator, copasi, ginsim, gna} to help the biologists define models of \emph{executable biology}~\cite{ex-bio}.
ANIMO (Analysis of Networks with Interactive MOdeling~\cite{animo-ieee}) is one of such tools. Its primary objective is to let the expert biologists
work directly on the formalization of their knowledge, supporting the generation of new insights on the studied processes~\cite{animo-gene}.
An ANIMO model is formed by two main parts: the network topology and the parameters.
The topology describes which biological components are included in the model, and which are the
interactions we want to represent. The parameters define the rate of occurrence of such interactions,
which are described based on simplified kinetic formulae.

Proteins are normally expressed at different concentrations in different individuals of the same species, and yet the
overall behavior of their biological networks does not differ significantly. This phenomenon has led to the notion that biological
networks are inherently \emph{robust}~\cite{bio-robust1, bio-robust2}. In modeling terms, this means that,
if a model is a close representation of a biological network, most of the parameters of that model
can vary inside a certain interval without influencing the qualitative behavior of the whole network.
The interactive approach of ANIMO is based on the assumption of robustness, as our tool is mainly aimed at the
development of network models with a focus on the topology.
Ideally, the biologist can ``play'' with the topology of a network, working
towards matching the qualitative behavior of experimental data, rather than precisely reproducing it.
However, it is not our intention to concentrate exclusively on the network topology:
in many cases a better parameter choice can improve the behavior of a network more than the addition of new components.
Indeed, making an unnecessarily complex model could reduce its usefulness both in terms of analysis
performances and closeness to reality. The first problem is simply due to the complexity of a network,
which would require more and more computational resources to be analysed\footnote{We also refer to the
problem of \emph{state space explosion}: when a model contains too many loosely coupled components,
the set of its possible evolutions grows exponentially, to the point of making it impossible to effectively apply analysis
techniques.}.
Realism of network models is more related to their
ultimate usefulness: a model that explains a particular behavior very well but contains many nodes marked
as ``unknown'' has little applicability, as its connection with known processes is very loose.
Therefore, it is desirable that a better parameter set is regularly sought for during the design cycle of
a complex biological network model. Some support for parameter choice was already provided in the first versions of ANIMO.
We present here an extended set of tools aimed at achieving a closer fit between ANIMO models and experimental data.
A guideline on how to use these tools to get the best results will also be presented as an ideal workflow.
Thanks to the better awareness on parameter choice gained through this new extension of ANIMO, the biologists
will be able to judge more easily which are the most promising topologies for a network, and thus drive
the experimental research more efficiently.

\section{ANIMO models}
\begin{figure}[htbp]
\begin{center}
\subfloat[\label{fig:animo-ui}]{\includegraphics[width=.55\textwidth]{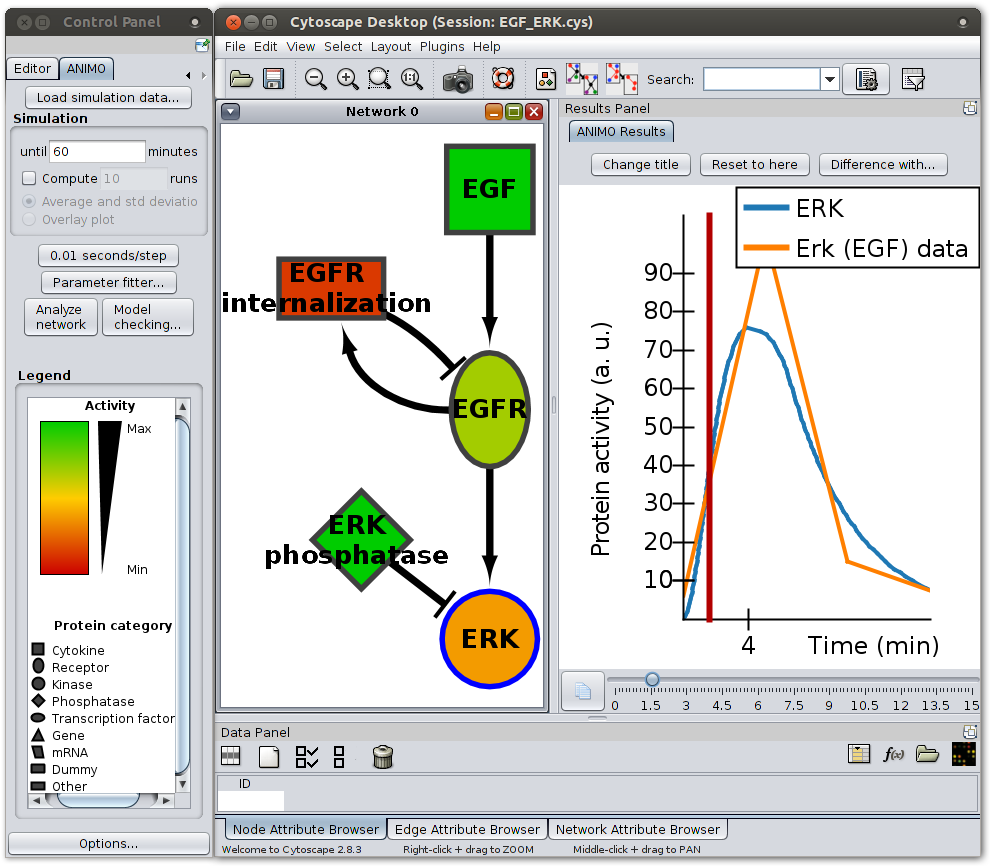}} \quad
\subfloat[\label{fig:animo-edge-ui}]{\includegraphics[width=.4\textwidth]{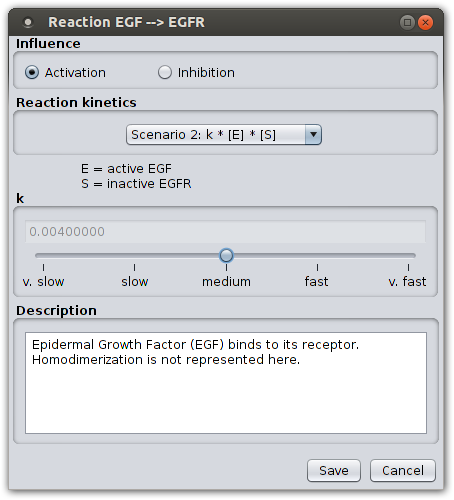}}
\caption{ANIMO user interface. (a) The main window of Cytoscape, with the ANIMO
plug-in showing a network model and the result of a simulation. (b) Editing the parameters for an ANIMO interaction.}
\end{center}
\end{figure}

The starting point of ANIMO is the traditional static representation of biological networks,
which can be easily drawn and managed in softwares like Cytoscape~\cite{cytoscape}. Indeed, ANIMO was implemented
as a plug-in to Cytoscape, with the aim of adding dynamics to the static representation of biological
networks, and thus allow for analysis on the behavior of such networks. The user interface of ANIMO
can be seen in Figure~\ref{fig:animo-ui}: at the center is the Network panel, where the network
model is represented in the familiar nodes-edges form used in the domain of biology.
Models in ANIMO are \emph{activity-based}, in the sense that nodes have their activity as main
property, and interactions among nodes change the activity of their targets. The concept of activity
can be intended for example as a generic post-translational modification a molecule can undergo to change its
function. In the case of a kinase, the phosphorylated state is usually interpreted as active.
In the context of a gene network, the activity of a node standing for a gene represents its current
transcriptional status.

A basic type of analysis that can be performed with ANIMO is the generation of
a simulation run, which is presented to the user in the form of a graph (to the right). The activity
graphs generated by ANIMO show the variation in activity of selected nodes over the course of the simulation run.
In addition to that, a slider placed under each graph allows the user to color the nodes in the Network panel
depending on their activity level at any point during the course of the simulation: the Legend
panel on the left links colors to activity. ANIMO models are based on the formalism of \tas~\cite{timed-automata-alur-dill},
which is not directly shown to the user. Indeed, ANIMO was designed with the aim of enabling expert biologists
to formalize their knowledge without the need for additional mathematical training.
For the details on the Timed Automata model used in ANIMO, we refer the interested reader to~\cite{animo-ieee}.

Of the two main components of an ANIMO model, only the topology is immediately visible to the user in the
Network panel; the parameters are accessed by double clicking the arcs representing node-node interactions.
The dialog window that is shown for an interaction contains the details of the abstract reaction kinetic
describing the interaction, together with the current value of its parameter $k$ (see Fig.~\ref{fig:animo-edge-ui}).
The unique parameter associated to any interaction in ANIMO is used as a scale factor to make the modelled
reaction occur faster or slower. We also provide the user with pre-set values for
$k$, encouraging an initial qualitative assignment of reaction rates as ``slow'', ``fast'' and so forth.
This approach is based on the assumption that biological networks are inherently robust:
once an acceptable set of parameters is found, a more precise parameter search will generally have
little impact on the fitness of a model to a reference data set.
However, as can be seen in the small example in Figure~\ref{fig:small-example-feedback},
robustness does not imply that any parameter choice will do.
In that example, all parameters are initially set to \emph{medium}
(Fig.~\ref{fig:small-example-feedback}{\sf a}).
The peak in EGFR was obtained setting the value of $k$ for {\sf EGFR internalization}~$\dashv$~{\sf EGFR}
to \emph{fast} (Fig.~\ref{fig:small-example-feedback}{\sf b}). As this is not enough
to get a peak also for ERK, which remains constantly inactive, a lower value of $k$
for {\sf ERK phosphatase}~$\dashv$~{\sf ERK} can be tried. A first attempt with
\emph{slow} (Fig.~\ref{fig:small-example-feedback}{\sf c}) leads to a low peak,
which can be increased by further lowering that parameter to \emph{very slow}
(Fig.~\ref{fig:small-example-feedback}{\sf d}). Further adjustment of the parameters
and non-default scenario choices (explained later) leads to the graph shown in Figure~\ref{fig:animo-ui}.
Network motifs such as the feedback loop shown in Figure~\ref{fig:small-example-feedback}
make models more dependent on parameter settings.
As biological networks tend to heavily rely on cross-talk, network topologies can rapidly become
complex. Some manual parameter fitting is currently necessary for the more complex ANIMO models.
This work is sometimes slow and error-prone, taking away some of the user-friendliness
for which ANIMO aims.
We will show how ANIMO has been improved to make parameter choice easier for the user.

\def\graphHeight{0.19\textheight}
\begin{figure}
\centering
\begin{tabular}{ccccccc}%
\includegraphics[height=\graphHeight]{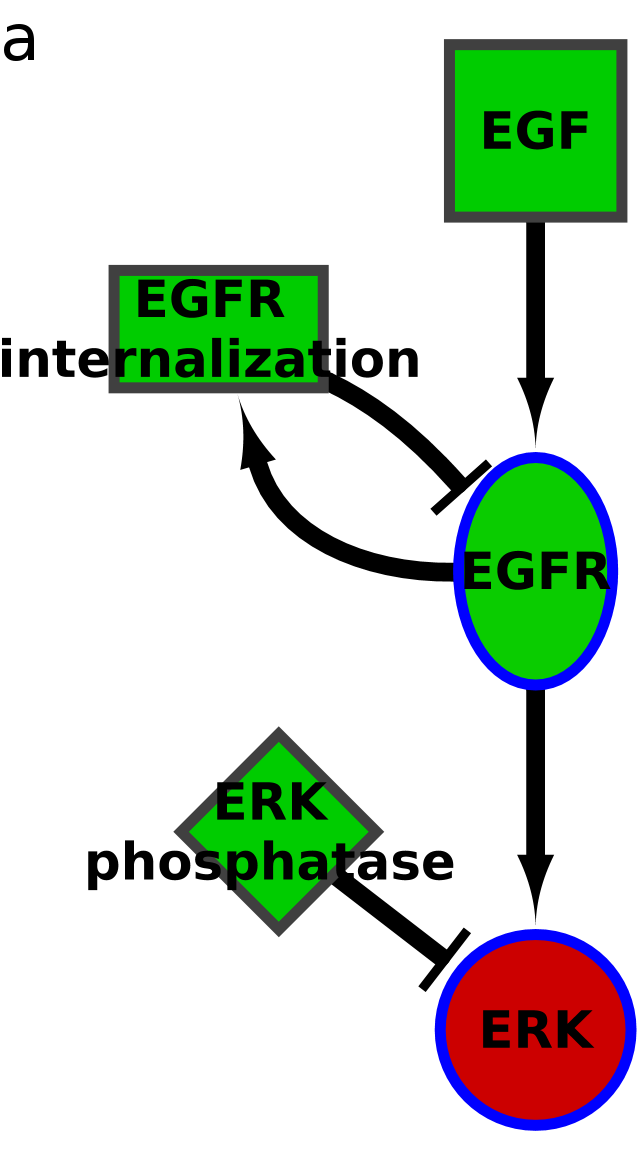} & \quad & \includegraphics[height=\graphHeight]{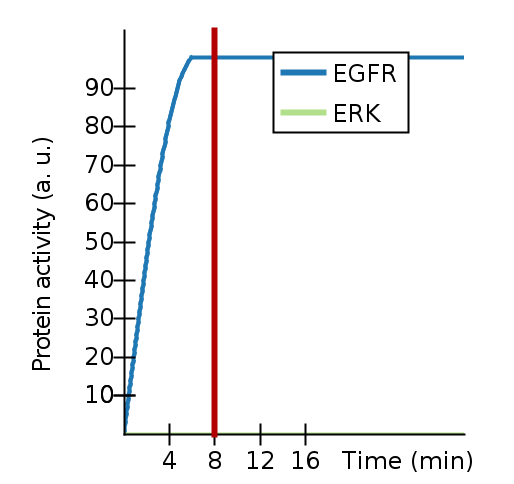} &\quad&
\includegraphics[height=\graphHeight]{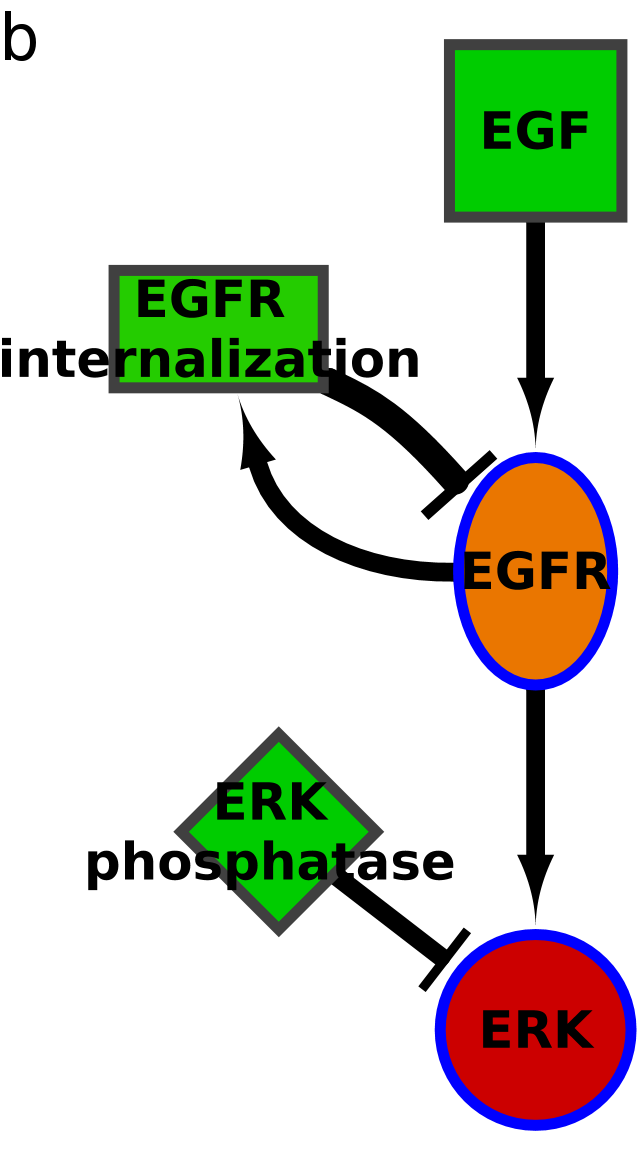} & \quad & \includegraphics[height=\graphHeight]{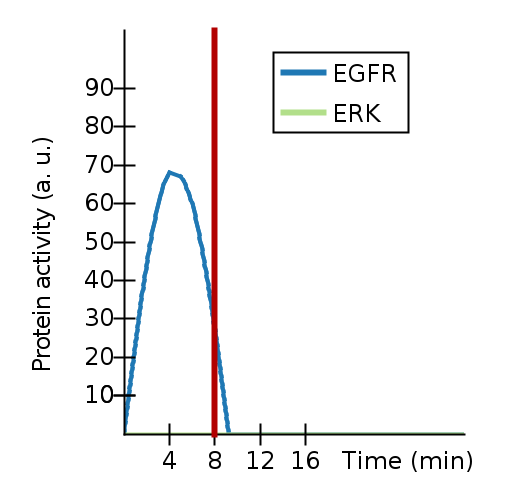} \\
\includegraphics[height=\graphHeight]{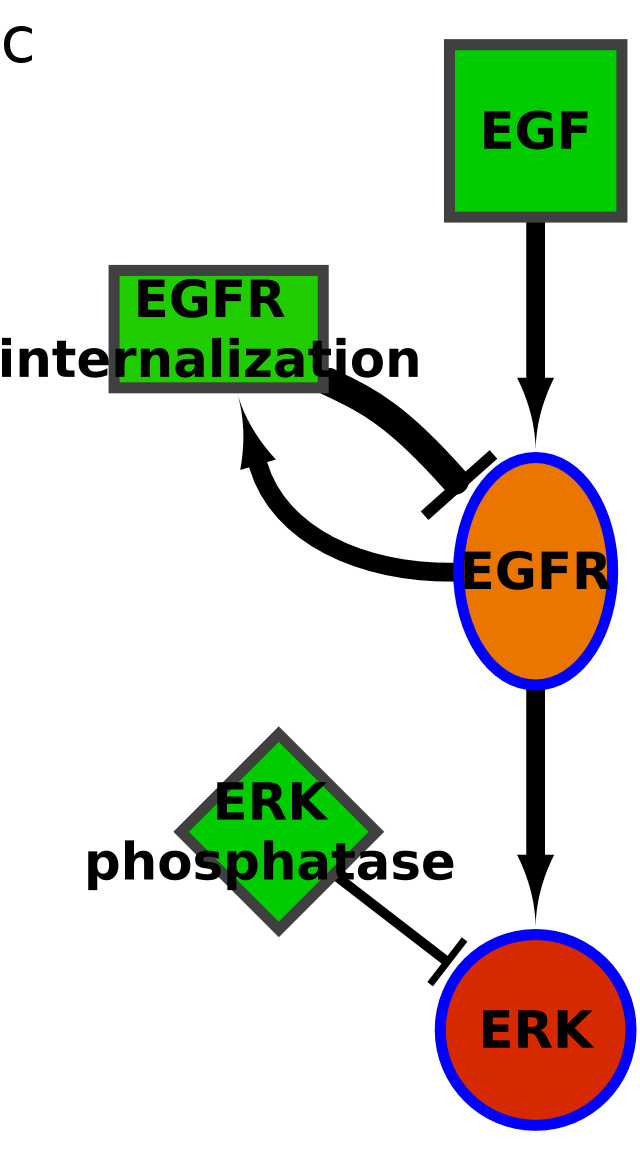} & \quad & \includegraphics[height=\graphHeight]{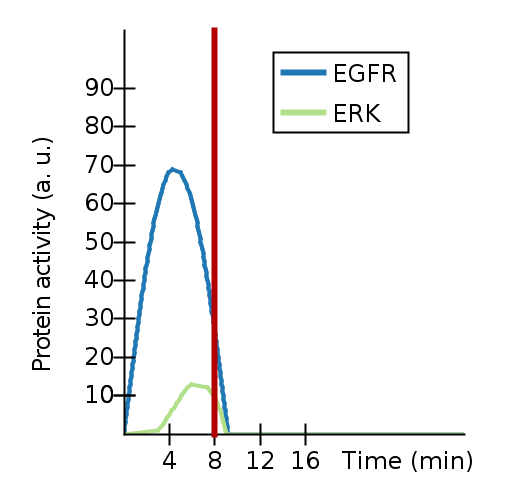} &\quad&
\includegraphics[height=\graphHeight]{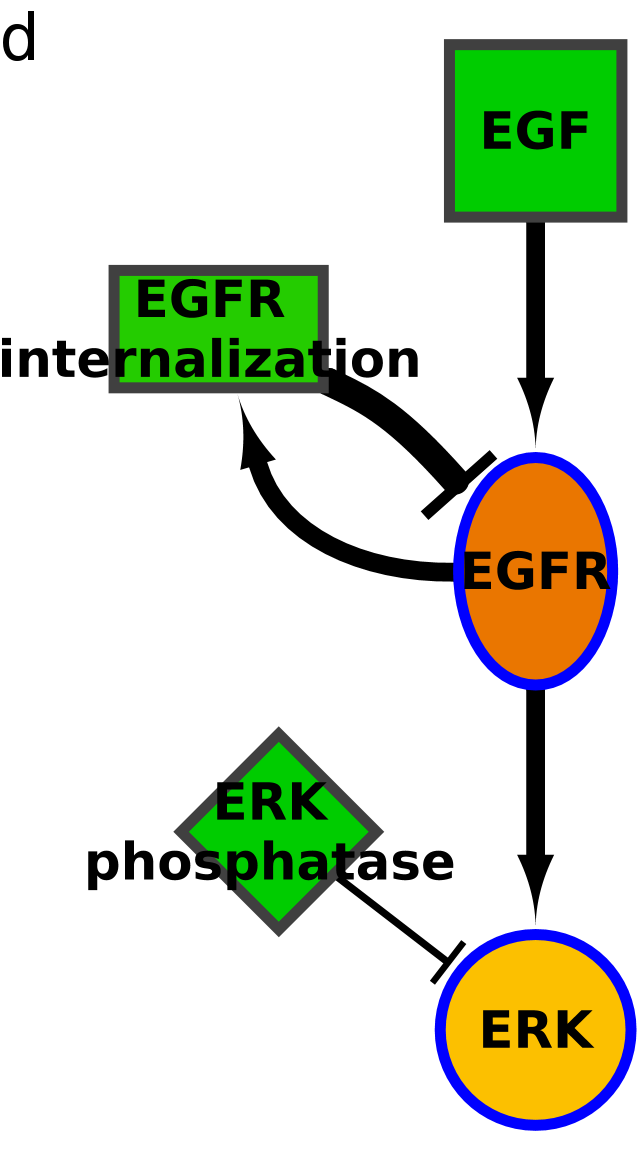} & \quad & \includegraphics[height=\graphHeight]{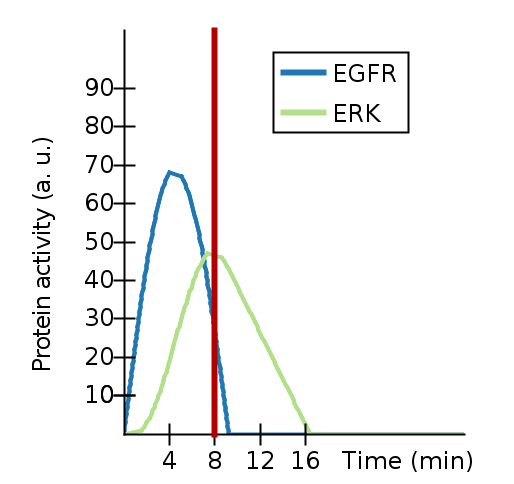} \\
\end{tabular}\ \\
\includegraphics[width=0.8\textwidth]{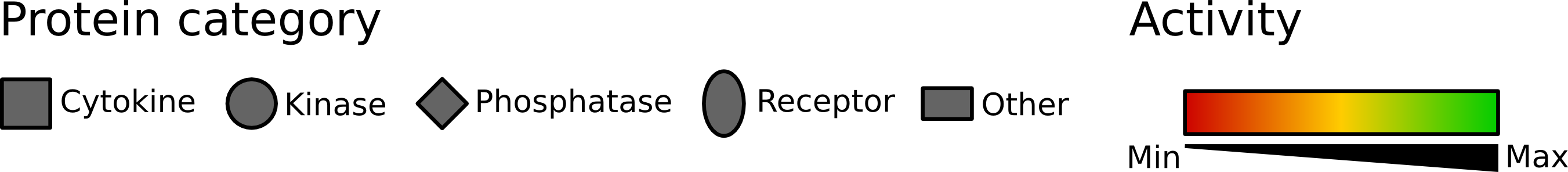}
\caption{Example parameter settings on a simple feedback network.
Arc width on the left represents parameter values: thicker arcs correspond to higher values.
Colors on the left represent node activities at the points in time highlighted by the vertical
red lines in the corresponding graphs on the right.\label{fig:small-example-feedback}}
\end{figure}

\section{Support for parameter synthesis in ANIMO}
When a model does not match experimental data, ANIMO offers three main tools to achieve a better fit:
manual parameter editing, comparisons between different model versions, and the newly introduced parameter sweep.

\subsection{Manual parameter editing}\label{sec:parameter-editing}
Double-clicking on an arrow in the Cytoscape representation of an interaction lets the user access
a dialog that allows to change the approximated scenario and parameters for the selected interaction.
The simplified scenarios were described in detail in~\cite{animo-ieee}, here a rough introduction is given.

In ANIMO models, the \emph{activity level} defines the fraction of a molecular
species that is considered active.
The activity level is expressed in ANIMO with an integer value between
0 (completely inactive) and a maximum value that represents complete activity. The maximum value can
be chosen by the user on a range between 1 and 100, determining the granularity (or precision)
with which the activity of the reactant is represented in the model. For example, an ANIMO node
with 2 levels of granularity (thus with maximum activity set to 1) can be used to represent a gene, which is considered as either
completely active or completely inactive. A more precise representation of the activity level
is necessary when we want to represent more complex dynamics, where also intermediate values are considered
important. The activity level of a node can be changed in an ANIMO model by way of activations
and inhibitions, which make it respectively rise and decrease. Each occurrence of an abstract
interaction changes the activity level of the downstream node by 1 step. For example,
consider the interaction MEK $\rightarrow$ ERK, which represents the activation of ERK by (active) MEK.
Each time that interaction occurs, the activity level of ERK rises by 1: if ERK has 100 activity levels,
it will take 100 interactions to take it from full inactivity to full activity. The rate
at which such interactions will occur is determined by the scenario and parameter of the
interaction.

The scenarios are used to choose which are the nodes to be taken into account
when computing the speed at which an interaction occurs.
\begin{description}
  \item[Scenario 1] the simplest of the three, this scenario approximates an interaction
    without taking into account the abundance of substrate (what is being
    changed by the interaction): the activation/inhibition rate
    depends only on the activity level of the upstream node. In the MEK $\rightarrow$ ERK
    example, the rate of the interaction would be linearly dependent on the current activity level of MEK.
  \item[Scenario 2] scenario 2 models the interaction rate
    to be directly dependent both on the activity
    level of the upstream node and on the availability of substrate. By substrate we
    mean the inactive downstream node in case of activation interaction, and active downstream node in case of inhibition.
    In the MEK $\rightarrow$ ERK example, the interaction rate is linearly dependent on
    both the activity level of MEK and the inactive fraction of ERK.
    The inactive level of a node is computed by subtracting the current activity level from the maximum activity.
  \item[Scenario 3] the nodes on which this scenario depends can be chosen directly by the user,
    and can be the active or inactive fraction of any two nodes in the network.
    This scenario can be used to represent an AND-gate, where two nodes are required to
    be simultaneously active in order to influence their target.
\end{description}

The choice for the parameter $k$ can be made directly by inserting a numeric value, or
indirectly by choosing a preset among the proposed qualitative values \emph{very slow, slow, medium, fast, very fast}.

\subsection{Comparing model versions}\label{sec:comparing-versions}
\begin{figure}
\centering
\includegraphics[width=.65\textwidth]{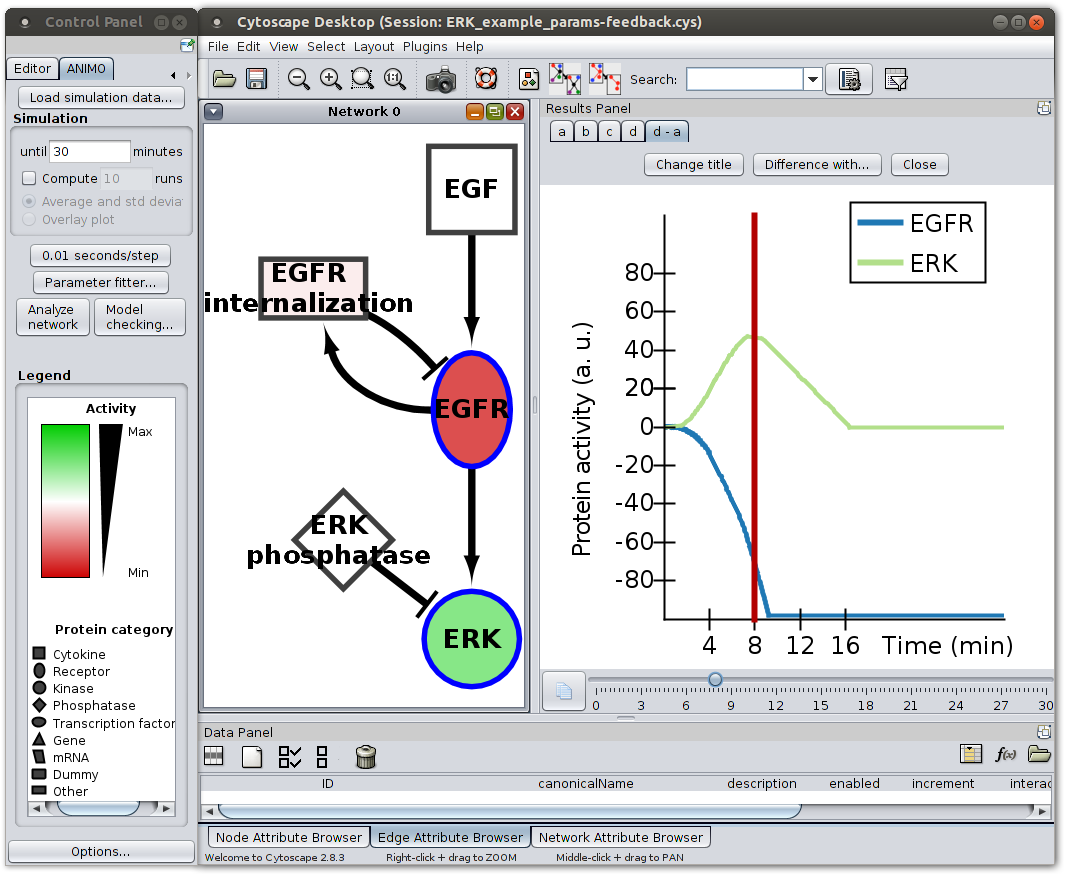}
\caption{Difference between the graphs in Figures~\ref{fig:small-example-feedback}{\sf a}
and~\ref{fig:small-example-feedback}{\sf d}. The Network panel shows the activity difference
in the network topology at the chosen simulation point represented by a vertical red line on the graph
in the Result panel.\label{fig:diff}}
\end{figure}

Particularly large networks make it more difficult for the user to understand the effects
of a change in network topology or interaction parameters. To overcome this difficulty, ANIMO
allows the user to visually compare two versions of a model in terms of their simulation results.
The Results panel in Figure~\ref{fig:animo-ui} shows the graph of selected node activities during
the course of an ANIMO simulation. The button labelled ``Difference with\dots'' allows to compare the current
simulation data with another based on a possibly different version of the model. When the two
simulations to be compared have been chosen, ANIMO produces a new graph plotting the difference
between the two original simulations. The slider under the new graph allows to visualize also
in the Network panel the changes in node activity in the whole network: Figure~\ref{fig:diff}
shows the difference between the first and last version of the model in Figure~\ref{fig:small-example-feedback}:
the difference was computed as {\sf d - a}. Nodes colored in green in the Network panel are
more active in version {\sf d}, while red nodes are more active in version {\sf a}.

By changing the title of a simulation, a user can keep track of different versions of their model,
which can be also kept and saved with the model, to be used in future sessions or for sharing purposes.
Should a user want to backtrack to a particular version of the network, the ``Reset to here'' button
can be used (see Fig.~\ref{fig:animo-ui}, buttons above the graph). As an additional help to recall what a model version consisted of, the tooltip of
the ``Reset to here'' button shows an image of the network topology, taken in the moment when
the selected simulation was begun. A click on the ``Reset to here'' button will reset the
network topology and parameters at the ones used to generate the selected simulation.
As the adaptation of a model may proceed on different paths, the ``Difference with\dots'' and
``Reset to here'' buttons provide the user with some help with tracking the changes to the model
and selecting the most promising ones.

\subsection{Parameter sweep}\label{sec:parameter-sweep}
In order to avoid introducing too many changes to the topology of a network, ANIMO users
should be reasonably sure that a topology does not fit a given data set before
modifying the structure of the network. As this would entail an extensive parameter search which,
even with the tools
described in Sections~\ref{sec:parameter-editing} and~\ref{sec:comparing-versions}, would
require a considerable amount of time and effort, we have provided the latest version of ANIMO
with a support for parameter sweeps. The idea is to let ANIMO automatically explore a user-defined parameter
space by computing a series of simulations based on a constant network topology, where only the
parameters are being changed. Comparing the results with a given experimental data series
will allow for an automatic pre-screening, leaving the user with a choice among the best
fitting settings.
Moreover, the task of parameter sweeping can be trivially parallelized, allowing us to
exploit the multi-core architectures available in most personal computers, and present the
users with the requested results in considerably less time.
In our case, the multi-core parameter sweep is implemented by using a thread pool pattern~\cite{thread-pool},
relying on the local scheduler to have the threads distributed on all CPU cores.

Another important option provided in ANIMO is the choice between linear and logarithmic scales
when defining the way a parameter space has to be explored: this gives the user more
control on the balance between precision and performances. When considering a big set of
interactions or of parameter settings, it is faster to proceed in logarithmic steps. After
a more precise interval of parameters has been identified for a subset of the interactions,
a linear search becomes more feasible. In this way, the computational resources can be used more efficiently
and results can be obtained faster.

The goodness of a model configuration is determined in ANIMO by comparing the activity graph
resulting from a simulation with the selected configuration
against a given experimental data set, applying the following formula for each of the selected data series $i$:
$$
\mbox{\it Error}_i = \max_{t = 0}^{\mbox{\it \scriptsize maxTime}} \frac{|\mbox{\it Data}_i(t) - \mbox{\it Model}_i(t)|}{\mbox{\it nLevels}_i}\\
$$
where $\mbox{\it Data}_i(t)$ is the experimental data point at time $t$ for the series $i$,
$\mbox{\it Model}_i(t)$ is the corresponding value on the graph computed by ANIMO
with the current parameter configuration, and $\mbox{\it nLevels}_i$
is the granularity (number of levels) of the network node $i$.
The error of the given model configuration is then computed as the maximum error over
all selected data series:
$$
\mbox{\it Error} = \max_{i = 0}^{\mbox{\it \scriptsize nSeries}}(\mbox{\it Error}_i)
$$

Thanks to the support for parameter sweeps, ANIMO users can obtain better insight on the
parameter sensitivity of their models, identifying critical points in the networks.
This can help identifying the most sensitive
areas of a network, and possibly suggest some intervention points for successive topology expansions,
if model robustness needs to be enforced. Note that the approach is only sketched here:
statistical considerations on confidence should also be taken into account.

All the simulations computed in ANIMO when performing a parameter sweep are based on
a deterministic version of the model: a given parameter configuration gives always the same
simulation as result. This is why we use only one simulation per configuration
when comparing the results with experimental data. However, the user can add some
non-determinism to the model by defining uncertainty intervals around which interaction
times are distributed. For example, the default setting of 5\% uncertainty illustrated in~\cite{animo-ieee}
implies that an interaction can complete in a time $t$ sampled from an uniform distribution
in the continuous interval $[0.95 \times T, 1.05 \times T]$.
Here, $T$ is the exact time for one interaction step, and is computed by applying
the selected scenario to the current activity levels of the nodes involved in the interaction.
Adding uncertainty to a model can be used to test its robustness. The user can then
ask ANIMO to compute a number of simulations in \emph{overlay} mode,
which will allow the user to see all the simulations superposed in one graph,
and realize whether in some cases the model significantly deviates from its normal behavior.

We now present an example application of ANIMO's parameter sweep feature on the model shown in Figure~\ref{fig:animo-ui}.
Thanks to parameter sweep, we can make a more exhaustive search than the one presented in Figure~\ref{fig:small-example-feedback},
using an automatic approach instead of a manual trial-and-error process.
We started by choosing scenario 2 for the two interactions involving ERK, while the other three interactions were left to scenario 1:
this allows us to obtain a smoother looking graph for ERK.
A logarithmic parameter sweep was performed, letting the parameter values
vary over the 5 pre-sets ``very slow'' ($k = 0.001$), ``slow'' ($k = 0.002$), ``medium'' ($k = 0.004$), ``fast'' ($k = 0.008$), ``very fast'' ($k = 0.016$)
for all 5 interactions in the network. This resulted in $\mbox{\it parameter choices}^{\mbox{\it\scriptsize interactions}} = 5 ^ 5 = 3\,125$ simulations, that were computed
in 43 seconds 
on an eight-core Intel\circledR\ Core$^{\mbox{\scriptsize\texttrademark}}$ i7 CPU at 2.80GHz equipped with 4 Gb RAM
and running Ubuntu GNU/Linux 13.10 64bit. The model we consider here is small enough for one simulation to take
less than a second, but the exponential growth of the number of simulations required for a parameter sweep
can make the sheer number of tasks challenging for any processing unit.
For example, a more complex model involving 10 interactions instead of 5 would require $5^{10} = 9\,765\,625$ simulations,
which could be computed in approximately a day and a half on an eight-core machine.
Still, the possibility to compute simulations in parallel provides a significant help in reducing
the impact of an exponential growth of the space to be explored.

\begin{figure}
\centering
\includegraphics[width=0.92\textwidth]{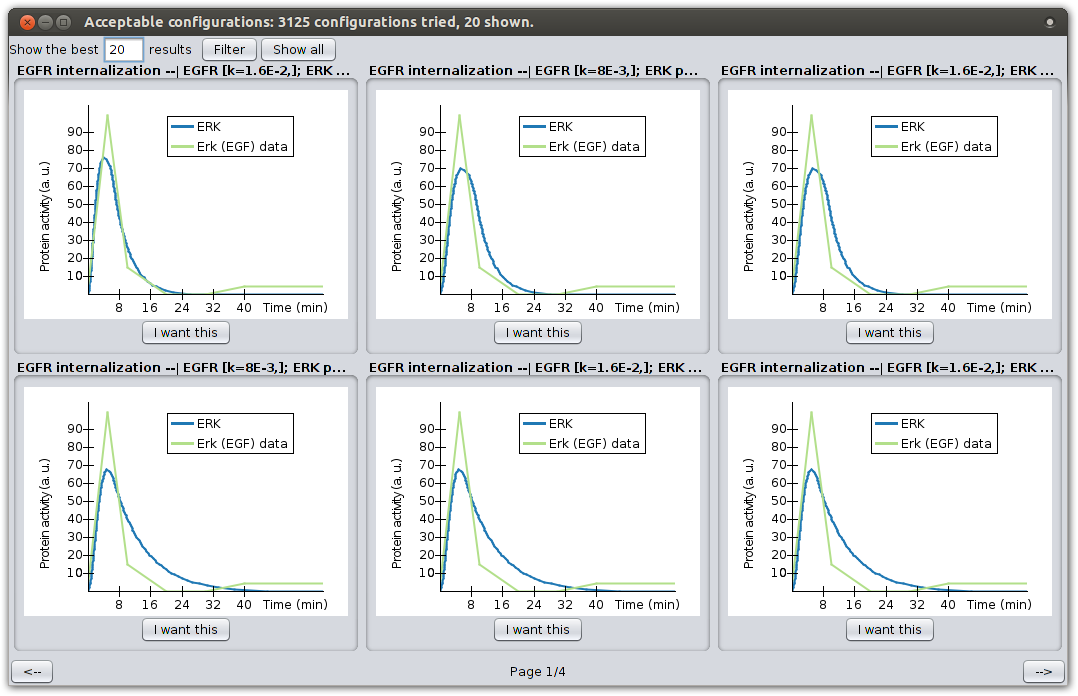}
\caption{Result window of a logarithmic parameter sweep in the ANIMO model of Figure~\ref{fig:animo-ui}.\label{fig:parameter-sweep-result}}
\end{figure}

The result window shown by ANIMO at the end of the computation of the $3\,125$ simulations can
be seen in Figure~\ref{fig:parameter-sweep-result}, where the best 6 results are listed in form of graphs.
By interacting with the interface, the user can choose to see more of the top-scoring results: this will allow to check
the width of the parameter settings for which the wanted behavior can still be considered ``close enough'' to
the experimental data. Pressing one of the buttons labelled ``I want this'' under one of the graphs
allows the user to copy the parameters used in the selected simulation to the network model.
The graph in Figure~\ref{fig:animo-ui} was obtained by choosing the first graph shown in Figure~\ref{fig:parameter-sweep-result},
i.e. setting the parameters as in Table~\ref{tab:parameters}. It can be seen that the model does not
fit the data very precisely: this is because some nodes known to influence the activity of ERK
were purposely left out of the example for simplicity's sake.


\begin{table}
\centering
\small
\begin{tabular}{|l|l|}
\hline
 Reaction & $k$ \\
\hline
\hline
EGF $\rightarrow$ EGFR & Fast \\
EGFR $\rightarrow$ EGFR internalization & Medium \\
EGFR internalization $\dashv$ EGFR & Very Fast \\
EGFR $\rightarrow$ ERK & Fast \\
ERK phosphatase $\dashv$ ERK & Medium\\
\hline
\end{tabular}
\caption{Qualitative values of the parameters $k$ used in the example model to obtain the graph shown in Figure~\ref{fig:animo-ui}.\label{tab:parameters}}
\end{table}

\section{Suggested ANIMO workflow}
In order to better profit from what ANIMO offers to the biologist, we suggest a step-by-step modeling workflow:
\begin{enumerate}
  \item\label{step:def-topology} Define a starting topology, based on literature data and current knowledge on the network.
  \item\label{step:set-rates} Choose the interaction parameters from the pre-set qualitative values,
    based on experience (e.g., protein expression is much slower than phosphorylation).
    The choice on the approximated kinetics can be for the most part based on scenario 1, referring to Section~\ref{sec:parameter-editing}.
  \item\label{step:check-behavior} Check that the network behaves as expected (e.g. addition of EGF makes ERK activity rapidly increase and successively decrease).
  \item Possibly iterate the steps~\ref{step:def-topology} to~\ref{step:check-behavior} until the network behaves as expected.
  \item Compare the model with experimental data.
  \item\label{step:precise-fit} If the model does not fit the data, change the parameters using more precise numerical values and/or scenario settings.
  \item\label{step:param-sweep} If the data fitting is still not satisfactory, apply parameter sweep to (subsets of) the model as described in Section~\ref{sec:parameter-sweep}.
  \item When a good fit is found, add non-determinism to the model and ensure that the
     behavior does not change significantly under reasonably high uncertainty settings.
  \item If the model can fit the data only with very precise parameter settings, or a
    parameter configuration that fits the data cannot be found after a reasonably extensive search
    on the parameter space, change the network topology possibly adding more nodes from literature.
    If no candidates can be found in literature, speculative candidates can be found by ``playing'' with the current
    configuration in ANIMO's user interface.
    We advise to let expert biologists perform this step on their own, or to perform it under their supervision,
    in order to keep the network both simple and realistic. For the same reason, we also advise not to introduce more than a couple of nodes at a time.
  \item Parameters for newly introduced interactions can be chosen as in step~\ref{step:set-rates}, possibly iterating
    steps~\ref{step:set-rates} to~\ref{step:param-sweep} to refine the parameters.
  \item To check how changes in the network are reflected in the behavior of the model, we advise to use the comparison tools
    described in Section~\ref{sec:comparing-versions}.
\end{enumerate}
If the model fits the data reasonably well only after modifications to the initial topology, it may contain some hypothetical or unexpected nodes.
In that case, additional laboratory experiments can be designed starting from the hypotheses introduced in the model, and
help the biologists decide which are the most precise explanations for the experimental observations.

\begin{figure}
\centering
\includegraphics[width=0.9\textwidth]{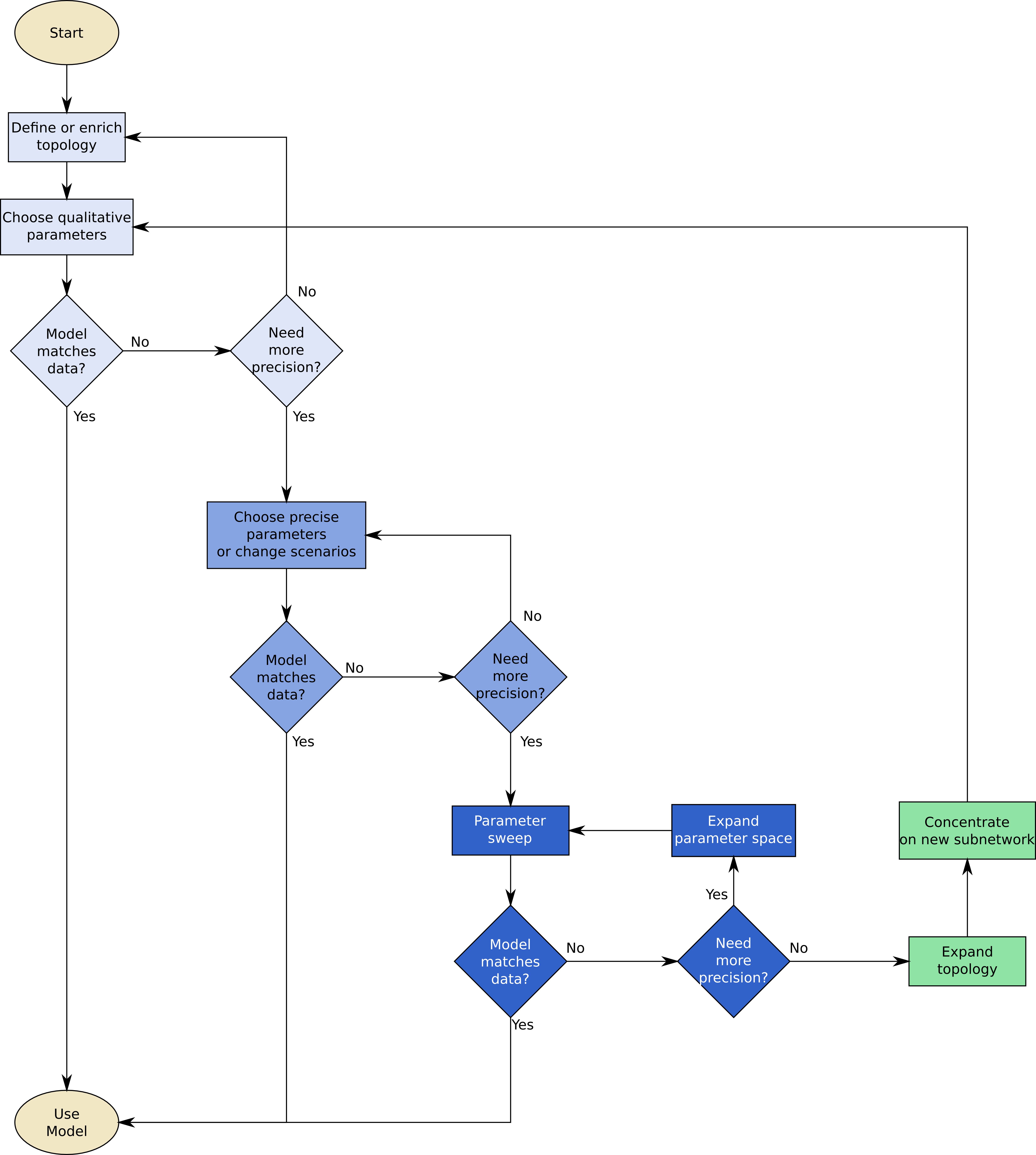}
\caption{ANIMO modeling workflow.\label{fig:workflow}}
\end{figure}

A simplified graphical representation of the workflow can be found in Figure~\ref{fig:workflow}: the three
degrees of precision with the parameter settings are highlighted with progressively darker shades of blue.
Moreover, the figure helps to stress the fact that in our method changes in topology (green boxes) should be made as a
last step, after a reasonable attempt at making the model fit via parameter adjustments.
Again, it is important to remember that if a model behaves correctly only for very narrow parameter settings
then it may not be as robust as expected, so a review of its topology may be needed.

The process described above allows the biologists to formalize their knowledge,
and makes it easier to have group discussions when trying to combine multiple subnetworks in a more comprehensive model.
It is worth noting that experience in both the specific biological setting and generic modeling of biological
events represents an important asset for a faster and more realistic modeling process with ANIMO. However, as
we claimed with all previous versions of the tool, it is still the case that no experience in formal methods
is needed in order to use ANIMO. In particular, an ideal ANIMO user could have no experience on the \tas\ foundations
of the tool and yet fully profit from its features.

\section{Conclusions}
ANIMO has been extended with a proper support for parameter synthesis. Now the tool can be used more efficiently
to model biological networks, without concentrating exclusively on their topology.
In particular, the addition of an automatic parameter sweep feature allows the user
to save considerable amounts of time, keeping closer to experimental data without the need
to perform a parameter search by hand.

In the future, we plan to improve our parameter analysis techniques, possibly using some of
the techniques already developed for testing parameter sensitivity~\cite{inverse-ta1, inverse-ta2} and
robustness~\cite{robust-ta1, robust-ta2, robust-ta3, robust-ta4} in \tas\ and hybrid systems more in general~\cite{decidability-ta-ha, imitator2}.
Techniques adopted in other formalisms such as e.g. differential equations~\cite{copasi, breach, kinfer}, rule-based languages~\cite{biocham, rulebender}
or Bayesian networks~\cite{bayes-infer} are also being
considered as a source of inspiration.
As already stated, parameter sensitivity is very important in biological models: if a treatment
derived from a model is based on narrow bounds for a parameter, that parameter must not exhibit significant
variety among patients. By introducing an automatic analysis for parameter sensitivity, our tool
could point out the zones of a network model where the user should concentrate more, for example by
improving the topology, or by performing more focused parameter sweeps.
Moreover, we also aim at applying some results from automata learning~\cite{test-based-modelling} to the generation of
\tas\ models based on experimental data: generating models that exhibit specific behaviors
implies a careful definition of their parameter space.
Finally, parametric model checking~\cite{param-mc} represents an interesting development direction:
having a tool automatically output the parameter intervals for which some
given outputs are observed is a highly valued resource for the biological researcher.
\bibliographystyle{eptcs}
\bibliography{Paper}
\end{document}